# Formal and Informal Model Selection with Incomplete Data

**Geert Verbeke, Geert Molenberghs and Caroline Beunckens**

*Abstract.* Model selection and assessment with incomplete data pose challenges in addition to the ones encountered with complete data. There are two main reasons for this. First, many models describe characteristics of the complete data, in spite of the fact that only an incomplete subset is observed. Direct comparison between model and data is then less than straightforward. Second, many commonly used models are more sensitive to assumptions than in the complete-data situation and some of their properties vanish when they are fitted to incomplete, unbalanced data. These and other issues are brought forward using two key examples, one of a continuous and one of a categorical nature. We argue that model assessment ought to consist of two parts: (i) assessment of a model's fit to the observed data and (ii) assessment of the sensitivity of inferences to unverifiable assumptions, that is, to how a model described the unobserved data given the observed ones.

## 1. INTRODUCTION

In many longitudinal and multivariate settings, not all designed measurements are collected. The implications of incompleteness need to be carefully considered and incorporated in the modeling process. Early work was largely concerned with algorithmic and computational solutions to the induced lack of balance or other design deviations (Afifi and Elashoff, 1966; Hartley and Hocking, 1971). Nowadays, general algorithms such as expectation-maximization (EM, Dempster, Laird and Rubin, 1977), and data imputation and augmentation procedures (Rubin, 1987), combined with powerful computing resources and flexible software implementations, are available. Thus, emphasis should be on assessing the impact of missing data on subsequent statistical inference.

We use terminology of Little and Rubin (2002, Chapter 6). A nonresponse process is *missing completely at random* (MCAR) if missingness is independent of unobserved and observed data and *missing at random* (MAR) if, conditional on the observed data, missingness is independent of the unobserved measurements. A process that is neither MCAR nor MAR is termed *nonrandom* (MNAR).

Given MAR, a valid analysis ignoring the missing-value mechanism can be obtained, within a likelihood or Bayesian framework, provided the parameters describing the measurement process are functionally independent of those describing the missingness process. This is termed *ignorability* (Rubin,

*Geert Verbeke is Professor, Biostatistical Centre, Katholieke Universiteit Leuven, Kapucijnenvoer 35, B3000 Leuven, Belgium e-mail: geert.verbeke@med.kuleuven.de. Geert Molenberghs is Professor, Center for Statistics, Hasselt University, Agoralaan 1, B3590 Diepenbeek, Belgium e-mail: geert.molenberghs@uhasselt.be. Caroline Beunckens is Postdoctoral Research Fellow, Center for Statistics, Hasselt University, Agoralaan 1, B3590 Diepenbeek, Belgium e-mail: caroline.beunckens@uhasselt.be.*

1976, Little and Rubin, 2002) and simplifies modeling (Diggle, 1989; Verbeke and Molenberghs, 2000). Such *direct-likelihood* and direct Bayesian analyses are increasingly preferred over ad hoc methods such as *last observation carried forward* (LOCF), *complete case analysis* (CC) or single imputation (Molenberghs et al., 2004; Mallinckrodt et al., 2003a, 2003b; Jansen et al., 2006a). Practically, tools like the linear and generalized linear mixed-effects models (Verbeke and Molenberghs, 2000; Molenberghs and Verbeke, 2005) can be used.

Nevertheless, in spite of the flexibility and elegance a direct-likelihood method brings, there are fundamental issues when selecting a model and assessing its fit to the observed data, which do not occur with complete data. These are the central theme of this paper. The issues discussed occur already in the MAR case, but they are compounded further under MNAR. One can never fully rule out MNAR. However, it is usually difficult to justify the particular choice of MNAR model (Jansen et al., 2006b). For example, different MNAR models may fit the observed data equally well, but engender quite different implications for the unobserved measurements and conclusions drawn. Without additional information, one can only distinguish between such models using their fit to the observed data, and so goodness-of-fit tools alone do not provide a relevant means of choosing between such models, naturally leading to sensitivity analysis, broadly defined as an instrument to assess the impact on statistical inferences from varying the, often untestable, assumptions in an MNAR model (Vach and Blettner, 1995; Copas and Li, 1997; Scharfstein, Rotnitzky, and Robins, 1999; Molenberghs and Kenward, 2007).

The ideas will be developed by means of two running examples with their model families, introduced in Section 2, along with initial analyses. Issues arising when analyzing incomplete data, under MAR as well as MNAR, are listed in Section 3. Ways of tackling the problems are the subject of Section 4.

## 2. RUNNING EXAMPLES AND THEIR INITIAL ANALYSES

### 2.1 The Orthodontic Growth Data

For 11 girls and 16 boys, the distance from the center of the pituitary to the maxillary fissure was recorded at ages 8, 10, 12 and 14 (Pothoff and Roy, 1964). Little and Rubin (2002) deleted 9 of the $[(11+16)\times 4]$ observations, thereby producing 9 incomplete subjects with a missing measurement at age 10. Their missingness generating mechanism was such that subjects with a low value at age 8 are more likely to have a missing value at age 10. Data tabulations and graphical displays can be found in Verbeke and Molenberghs (2000) and Molenberghs and Kenward, (2007).

Jennrich and Schluchter (1986), Little and Rubin (2002) and Verbeke and Molenberghs (1997, 2000) fitted eight linear mixed models, of the form $\mathbf{Y}_i = X_i\boldsymbol{\beta} + Z_i\mathbf{b}_i + \boldsymbol{\varepsilon}_i$, where $\mathbf{Y}_i$ is the $(4\times 1)$ response vector, $X_i$ is a $(4\times p)$ design matrix for the fixed effects, $\boldsymbol{\beta}$ is a vector of unknown fixed regression coefficients, $Z_i$ is a $(4\times q)$ design matrix for the random effects, $\mathbf{b}_i$ is a zero-mean $(q\times 1)$ vector of normally distributed random parameters, with covariance matrix $D$, $\boldsymbol{\varepsilon}_i$ is a zero-mean normally distributed $(4\times 1)$ random error vector, with covariance matrix $\Sigma$, and $\mathbf{b}_i$ and $\boldsymbol{\varepsilon}_i$ are independent. The mean $X_i\boldsymbol{\beta}$ will be a function of age, sex, and/or the interaction between both.

Model 1 leaves the group by time model and the covariance matrix unstructured. Through model simplification steps, details of which can be found in Verbeke and Molenberghs (2000), passing via nonparallel (Model 2) and parallel (Model 3) straight mean profiles, Model 7 is retained, featuring nonparallel straight mean profiles and a compound-symmetry covariance structure. Little and Rubin (2002) fitted the same models to the trimmed, incomplete, version of the dataset, using direct-likelihood methods, and were led to the same Model 7. A quite different picture would emerge, were simple, ad hoc methods used (Molenberghs and Kenward, 2007). As is commonly known in the research community, analyses like last observation carried forward (LOCF), complete case analysis, and simple forms of mean imputation produce distorted mean and/or covariance structures. This message is in line with the unreliable performance of such simple methods, as opposed to direct likelihood, thanks to the latter method's validity under MAR. It is often argued that the price to pay is the need to fit a model to the entire longitudinal sequence through, for example, a linear mixed model, even in circumstances where scientific interest focuses on the last planned measurement occasion. However, for balanced longitudinal data, where the number of subjects is sufficiently large compared to the number of times, a full multivariate normal (Model 1) can often be considered,

TABLE 1
*The orthodontic growth data*

| Principle | Method | Boys at age 8 | Boys at age 10 |
|---|---|---|---|
| Original | ML | 22.88 (0.56) | 23.81 (0.49) |
| | REML ≡ MANOVA | 22.88 (0.58) | 23.81 (0.51) |
| | ANOVA per time | 22.88 (0.61) | 23.81 (0.53) |
| Observed | ML | 22.88 (0.56) | 23.17 (0.68) |
| | REML | 22.88 (0.58) | 23.17 (0.71) |
| | MANOVA | 24.00 (0.48) | 24.14 (0.66) |
| | ANOVA per time | 22.88 (0.61) | 24.14 (0.74) |
| CC | ML | 24.00 (0.45) | 24.14 (0.62) |
| | REML ≡ MANOVA | 24.00 (0.48) | 24.14 (0.66) |
| | ANOVA per time | 24.00 (0.51) | 24.14 (0.74) |
| LOCF | ML | 22.88 (0.56) | 22.97 (0.65) |
| | REML ≡ MANOVA | 22.88 (0.58) | 22.97 (0.68) |
| | ANOVA per time | 22.88 (0.61) | 22.97 (0.72) |

Likelihood, MANOVA and ANOVA analyses for the original data and the trimmed data (observed, CC and LOCF). Means for boys at ages 8 and 10 are displayed.

not making assumptions beyond the ones made by, say, multivariate analysis of variance (MANOVA), ANOVA per time point or, equivalently, a $t$ test per time point. This is illustrated in Table 1, using Model 1 fitted to the complete and trimmed growth data. Means for boys at the ages 8 and 10 are displayed. Whenever the data are balanced, the means are the same regardless of which estimation method is used. Standard errors are asymptotically equal, and even in our small sample differences are negligible. CC overestimates the means since the subjects removed from analysis have lower-than-average means, and LOCF underestimates the mean at age 10, since the age-8 measurement is carried forward.

Analyzing the trimmed data, the results from the direct-likelihood analyses, valid under MAR, diverge from the frequentist MANOVA and ANOVA analyses, the latter valid only under MCAR. MANOVA effectively reduces to CC, owing to its inability to take incomplete sequences into account. ANOVA produces correct inferences only at occasions with complete data.

## 2.2 The Slovenian Public Opinion Survey

In 1991 Slovenians voted for independence from former Yugoslavia in a plebiscite. To prepare for this result, the Slovenian government collected data in the Slovenian Public Opinion Survey (SPO), a month prior to the plebiscite. Rubin, Stern and Vehovar (1995) studied the three fundamental questions, for the one time added to the usual SPO questions and, in comparing it to the plebiscite's outcome, drew conclusions about the missing data process. Molenberghs, Kenward and Goetghebeur (2001) used these data to introduce sensitivity analysis methodology. Details can be found in Molenberghs and Verbeke (2005) and Molenberghs and Kenward (2007). The three questions were: (1) Are you in favor of Slovenian independence? (2) Are you in favor of Slovenia's secession from Yugoslavia? (3) Will you attend the plebiscite? Question (3) is relevant due to the political decision that not attending was treated as an effective NO to question (1). The primary estimand, the proportion $\theta$ of people that will be considered as voting YES, follows from those answering yes to both the attendance and independence questions. An overview of various analyses is given in Molenberghs and Kenward (2007).

These authors used the model proposed by Baker, Rosenberger and DerSimonian (BRD, 1992) for two-way contingency tables subject to nonmonotone missingness. Organize the two outcomes, together with their missingness indicators, as a four-way contingency table with counts $Z_{r_1,r_2,jk}$, where $j,k=0,1$ reference the two categories for the response variables and $r_1,r_2=0,1$ refer to the two missingness indicators. The counts are not fully observable and should be seen as a device to facilitate modeling. Such modeling takes place in terms of cell probabilities $\nu_{r_1,r_2,jk}$ with the same indexing system as the counts $Z_{r_1,r_2,jk}$. Rewrite the probabilities governing the incomplete patterns as modified version of the complete-pattern probabilities $\nu_{11,jk}$, that is, $\nu_{10,jk}=\nu_{11,jk}\beta_{jk}$, $\nu_{01,jk}=\nu_{11,jk}\alpha_{jk}$ and $\nu_{00,jk}=\nu_{11,jk}\alpha_{jk}\beta_{jk}\gamma$. The $\alpha$ ($\beta$) parameters describe missingness in the independence (attendance) question, and $\gamma$ captures the interaction between both. BRD considered nine models, based on setting $\alpha_{jk}$ and $\beta_{jk}$ constant in one or more indices: BRD1: $(\alpha,\beta)$; BRD4: $(\alpha,\beta_k)$; BRD7: $(\alpha_k,\beta_k)$; BRD2: $(\alpha,\beta_j)$; BRD5: $(\alpha_j,\beta)$; BRD8: $(\alpha_j,\beta_k)$; BRD3: $(\alpha_k,\beta)$; BRD6: $(\alpha_j,\beta_j)$; BRD9: $(\alpha_k,\beta_j)$. Interpretation is straightforward; for example, BRD1 is MCAR, and in BRD4 missingness in the first variable is constant, while missingness in the second variable depends on its value. BRD6–BRD9 saturate the observed data degrees of freedom; the lower-numbered ones do not, leaving room for nontrivial fit to the observed data.

Rubin, Stern and Vehovar (1995) conducted several analyses of the data. Their main emphasis was on determining the proportion $\theta$ of the population

Table 2
*The Slovenian Public Opinion Survey*

| Model | Structure | d.f. | loglik | $\widehat{\theta}$ | C.I. | $\widehat{\theta}_{\mathrm{MAR}}$ |
|---|---|---|---|---|---|---|
| BRD1 | $(\alpha, \beta)$ | 6 | −2495.29 | 0.892 | [0.878; 0.906] | 0.8920 |
| BRD2 | $(\alpha, \beta_j)$ | 7 | −2467.43 | 0.884 | [0.869; 0.900] | 0.8915 |
| BRD3 | $(\alpha_k, \beta)$ | 7 | −2463.10 | 0.881 | [0.866; 0.897] | 0.8915 |
| BRD4 | $(\alpha, \beta_k)$ | 7 | −2467.43 | 0.765 | [0.674; 0.856] | 0.8915 |
| BRD5 | $(\alpha_j, \beta)$ | 7 | −2463.10 | 0.844 | [0.806; 0.882] | 0.8915 |
| BRD6 | $(\alpha_j, \beta_j)$ | 8 | −2431.06 | 0.819 | [0.788; 0.849] | 0.8919 |
| BRD7 | $(\alpha_k, \beta_k)$ | 8 | −2431.06 | 0.764 | [0.697; 0.832] | 0.8919 |
| BRD8 | $(\alpha_j, \beta_k)$ | 8 | −2431.06 | 0.741 | [0.657; 0.826] | 0.8919 |
| BRD9 | $(\alpha_k, \beta_j)$ | 8 | −2431.06 | 0.867 | [0.851; 0.884] | 0.8919 |
| Model 10 | $(\alpha_k, \beta_{jk})$ | 9 | −2431.06 | [0.762; 0.893] | [0.744; 0.907] | 0.8919 |
| Model 11 | $(\alpha_{jk}, \beta_j)$ | 9 | −2431.06 | [0.766; 0.883] | [0.715; 0.920] | 0.8919 |
| Model 12 | $(\alpha_{jk}, \beta_{jk})$ | 10 | −2431.06 | [0.694; 0.905] | — | 0.8919 |

Analysis, restricted to the independence and attendance questions. Summaries on each of the Models BRD1–BRD9 are presented. In addition, intervals of ignorance and intervals of uncertainty for the proportion $\theta$ (confidence interval) attending the plebiscite following from fitting.

that would attend the plebiscite and vote for independence. The three other combinations of these two binary outcomes would be treated as voting "no." Pessimistic/optimistic bounds are obtained by setting all incomplete data that can be considered a yes (no), as yes (no); they equal $[0.694; 0.905]$. A complete case analysis produces $\widehat{\theta} = 0.928$ and an available case analysis $\widehat{\theta} = 0.929$. It is noteworthy that both estimates fall outside the pessimistic–optimistic interval and should be disregarded, since these seemingly straightforward estimators do not take the decision to treat absences as no's into account and thus discard available information. MAR based on two questions leads to $\widehat{\theta} = 0.892$ and, using the middle question as auxiliary, $\widehat{\theta} = 0.883$ is found. In contrast, their MNAR analysis produces $\widehat{\theta} = 0.782$. The plebiscite value is $\theta = 0.885$. Rubin, Stern and Vehovar (1995) concluded, owing to the proximity of the MAR analysis to the plebiscite value, that MAR in this and similar cases is a plausible assumption.

Molenberghs, Kenward and Goetghebeur (2001) and Molenberghs et al. (2007) fitted the BRD models and Table 2 summarizes the results. BRD1 produces $\widehat{\theta} = 0.892$, exactly the same as the first MAR estimate obtained by Rubin, Stern and Vehovar (1995). This is because both models assume MAR and use information from the two main questions.

## 3. COMPLEXITY OF MODEL SELECTION AND ASSESSMENT WITH INCOMPLETE DATA

Model selection and assessment are well-established components of statistical analysis, whether in cross-sectional or correlated settings; they are surrounded by several strands of intuition. First, it is researchers' common understanding that "observed≃expected" for a well-fitting model, usually understood to imply that observed and fitted profiles ought to be sufficiently similar in a longitudinal study, observed and fitted counts in contingency tables, etc. Second, for the special case of samples from normal distributions, the estimators for the mean vector and the variance-covariance matrix are independent, in small and large samples alike. Third, in the same situation, the least squares and maximum likelihood estimators are identical for mean parameters and asymptotically equal for covariance parameters. Fourth, in a likelihood-based context, deviances and related information criteria are considered appropriate tools for model assessment. Fifth, saturated models are uniquely defined and at the top of the model hierarchy. For contingency tables, a saturated model exactly reproduces the observed counts.

While it has been reasonably well known that these five points hold for well-balanced designs and complete sets of data, their failure with incomplete data is perhaps not as much part of operational knowledge as it should be. Therefore, we find it useful to provide illustrations by means of the running examples and by general considerations.

### 3.1 The "Observed≃Expected" Relationship

Figure 1 shows the observed and fitted mean structures for Models 1, 2, 3 and 7, fitted to the complete and incomplete versions of the growth dataset, respectively. Observed and fitted means for Model 1 coincide in the complete, balanced case, but do not for the trimmed data. The discrepancy is seen for the mean at age 10, the only one for which there is missingness.

### 3.2 The Mean–Variance Relationship in a Normal Distribution

To gain insight into the effect of the covariance structure on the mean structure, consider variations to Model 1, fitted to boys at ages 8 and 10. Retain an unstructured group-by-age mean structure, and pair it with three residual covariance structures: Model 1: unstructured; Model 7b: CS; Model 8b: independence. Fit these models to complete and incomplete data. For the complete data, the choice of covariance structure is immaterial, but the choice is crucial when data are incomplete. While at age 8 the point estimate remains 22.88 in all three cases, it varies from 23.17 for Model 1, over 23.52 for Model 7b, to 24.14 for Model 8b; the latter coincides with and hence is as bad as CC at age 10.

### 3.3 The Least Squares–Maximum Likelihood Difference

The difference between ordinary least squares and maximum likelihood is an issue, different from but related to the previous two. Table 1 reiterates that the MLE and the frequentist OLS differ for the incomplete data. Let us illustrate this well-known result for a bivariate normal (Section 3.3.1) and a contingency table (Section 3.3.2).

3.3.1 *A bivariate normal population.* Consider a bivariate normal population:

$$(1)\qquad \begin{pmatrix} Y_{i1} \\ Y_{i2} \end{pmatrix} \sim N\left( \begin{pmatrix} \mu_1 \\ \mu_2 \end{pmatrix}, \quad \begin{pmatrix} \sigma_1^2 & \sigma_{12} \\ \sigma_{12} & \sigma_2^2 \end{pmatrix} \right),$$

from which $i = 1, \ldots, N$ subjects are sampled. Assume that $d$ subjects complete the study and $N - d$ drop out after the first measurement.

In a frequentist available case method, using OLS, the parameters in (1) are estimated using the available information (Little and Rubin, 2002, Verbeke and Molenberghs, 2000): $\mu_1$ and $\sigma_1^2$ are estimated using all $N$ subjects, whereas only the remaining $d$ contribute to the other three parameters. For the mean parameters, this produces $\widehat{\mu_1} = (\sum_{i=1}^N y_{i1})/N$ and $\widetilde{\mu_2} = (\sum_{i=1}^d y_{i2})/d$. Little and Rubin (2002) present an explicit expression for the MLE, starting from the conditional density of the second outcome given the first one: $Y_{i2}|y_{i1} \sim N(\beta_0 + \beta_1 y_{i1}, \sigma_{2|1}^2)$, producing the maximum likelihood estimator $\mu_2$:

$$(2)\qquad \widehat{\mu_2} = \frac{1}{N}\left\{ \sum_{i=1}^{d} y_{i2} + \sum_{i=d+1}^{N} [\overline{y}_2 + \widehat{\beta_1}(y_{i1} - \overline{y}_1)] \right\}.$$

Here, $\overline{y}_1$ is the mean of the measurements at the first occasion among the completers. Several observations can be made. The difference between ML and OLS estimators vanishes only under MCAR and/or when $Y_{i1}$ and $Y_{i2}$ are uncorrelated.

Turning to the orthodontic growth dataset (see Table 3), a correction like (2) applies to the age of 10. From Figure 1(b), it is clear that those remaining on study have larger measurements than those removed, hence the downward correction in the likelihood estimator. The likelihood even overcorrects in this case, owing to a small sample size, since the estimated correlation between the ages 8 and 10 is substantially larger than the correlation between ages 10 and 12. We return to these points in the next section. There are also important consequences for model checking, since the observed-expected relationship is no longer straightforward. We return to this in Section 4.

Additionally, the coefficient $\beta_1$ depends on the variance components and hence a misspecified variance structure may lead to bias in $\widehat{\mu_2}$. This underscores the breakdown of the independence of the mean and variance estimators.

TABLE 3
*The orthodontic growth data*

| Data | Mean | Covar | Boys at age 8 | Boys at age 10 |
|---|---|---|---|---|
| Complete | unstr. | unstr. | 22.88 | 23.81 |
| | unstr. | CS | 22.88 | 23.81 |
| | unstr. | simple | 22.88 | 23.81 |
| Incomplete | unstr. | unstr. | 22.88 | 23.17 |
| | unstr. | CS | 22.88 | 23.52 |
| | unstr. | simple | 22.88 | 24.14 |

Comparison of mean estimates for boys at ages 8 and 10, complete and incomplete data, using direct likelihood, an unstructured mean model, and various covariance models.

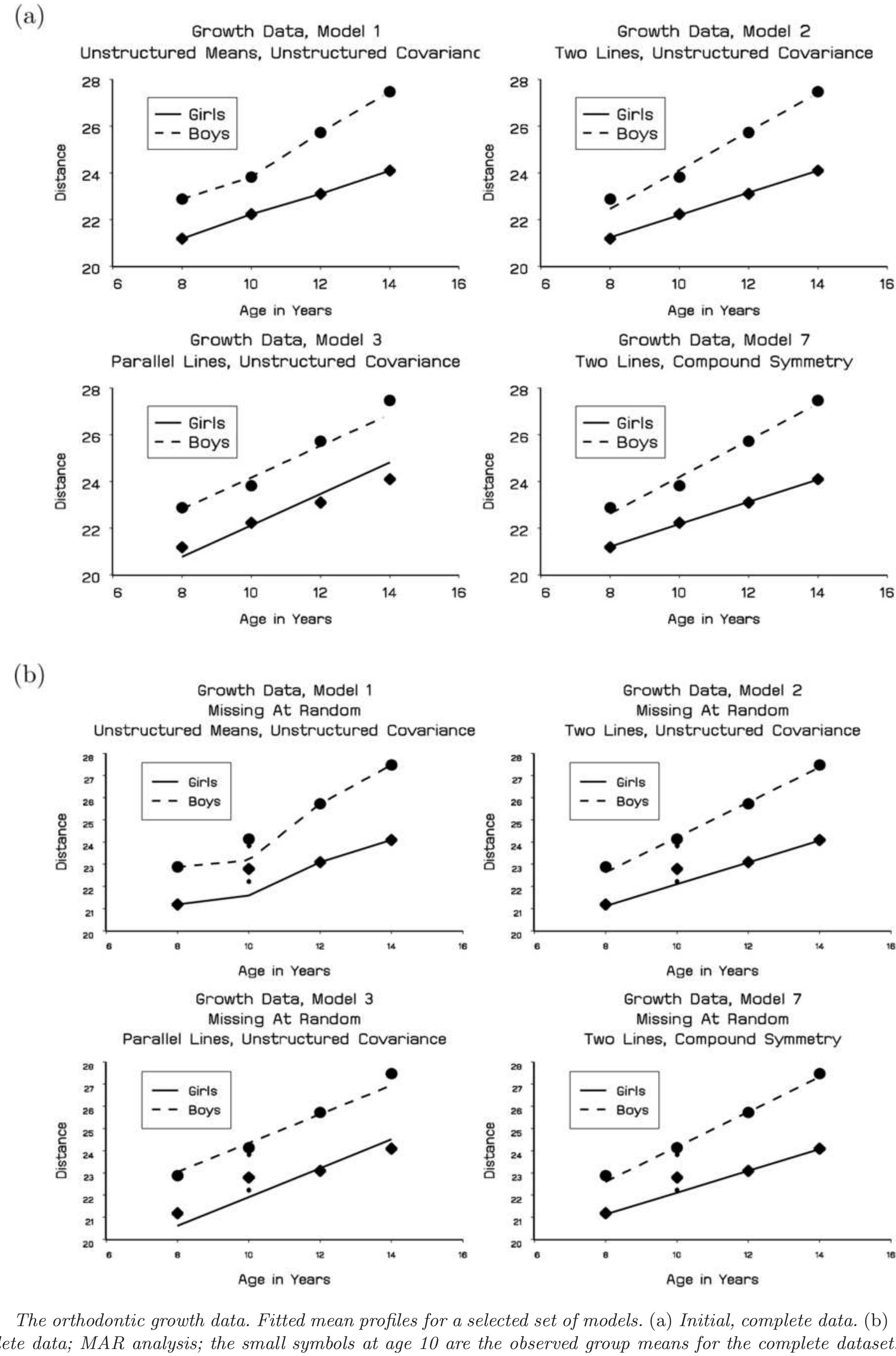

FIG. 1. *The orthodontic growth data. Fitted mean profiles for a selected set of models.* (a) *Initial, complete data.* (b) *Trimmed, incomplete data; MAR analysis; the small symbols at age 10 are the observed group means for the complete dataset.*

The adequate performance of Model 7b owes to the fact that the expected mean of a missing age-10 measurement gives equal weight to all surrounding measurement, rather than overweighting the age-8 measurement due to an accidentally high correlation. The zero correlations in Model 8b do not allow for such a correction, hence its coincidence with CC.

3.3.2 *An incomplete contingency table.* Consider an incomplete $2 \times 2$ contingency table:

$$\begin{array}{|c|c|}\hline Z_{1,11} & Z_{1,12} \\ \hline Z_{1,21} & Z_{1,22} \\ \hline \end{array} \qquad \begin{array}{|c|}\hline Z_{0,1} \\ \hline Z_{0,2} \\ \hline \end{array},$$

where $Z_{r,jk}$ refers to the number of subjects in the completers ($r = 1$) and dropouts ($r = 0$) groups, respectively, with response profile $(j, k)$. Since for the dropouts only the first outcome is observed, only summaries $Z_{r=0,j}$ are observable. Using all available data, the probability of success at the first time is estimated as $\widehat{\pi_1} = (Z_{1,1+} + Z_{0,1+})/N$, where subscript "+" refers to summing. Using available cases only, the estimator for the success probability at the second time is $\widetilde{\pi_2} = Z_{1,+1}/d$. Thus, OLS does not use information from incomplete subjects, whereas the MLE does:

$$\widehat{\pi_2} = (Z_{1,+1} + Z_{0,1} \cdot Z_{1,11}/Z_{1,1+} + Z_{0,2} \cdot Z_{1,21}/Z_{1,2+})/N. \tag{3}$$

### 3.4 Deviances and Saturated Models

Revisiting Table 2, a deviance comparison between BRD1 and any of BRD2–5 and of the latter with BRD6–9, shows the earlier models fit poorly. We are thus left with BRD6–9 as candidates. Now, all four produce exactly the same likelihood at maximum, owing to saturation. Nevertheless, the estimates for $\theta$ differ between these four models, since $\theta$ is a function, not only of the model fit to the observed data, but also of the model's *prediction* of the unobserved data, given what has been observed. Here, and in what follows, we will use "'fit" to indicate, broadly, the relationship between observed data and a model, with "prediction" reserved for the relationship between unobserved data and a model. The concepts of "fit/prediction" and "saturation" can be seen as relative to either the observed or the complete data, posing challenges for model selection and fit assessment. In particular, a model that saturates *all* degrees of freedom would by definition be nonidentifiable.

TABLE 4
*The Slovenian public opinion survey*

Observed data and fit of BRD7, BRD7(MAR), BRD9 and BRD9(MAR) to incomplete data

| | | | | | | | |
|---|---|---|---|---|---|---|---|
| 1439 | 78 | 159 | | 144 | 54 | 136 | |
| 16 | 16 | 32 | | | | | |

Prediction of BRD7 to complete data
≡ Completed data using BRD7 fit

| | | | | | | | |
|---|---|---|---|---|---|---|---|
| 1439 | 78 | 3.2 | 155.8 | 142.4 | 44.8 | 0.4 | 112.5 |
| 16 | 16 | 0.0 | 32.0 | 1.6 | 9.2 | 0.0 | 23.1 |

Prediction of BRD9 to complete data
≡ Completed data using BRD9 fit

| | | | | | | | |
|---|---|---|---|---|---|---|---|
| 1439 | 78 | 150.8 | 8.2 | 142.4 | 44.8 | 66.8 | 21.0 |
| 16 | 16 | 16.0 | 16.0 | 1.6 | 9.2 | 7.1 | 41.1 |

Prediction of BRD7(MAR) and BRD9(MAR) to complete data
≡ Completed data using BRD7(MAR) ≡ BRD9(MAR) fit

| | | | | | | | |
|---|---|---|---|---|---|---|---|
| 1439 | 78 | 148.1 | 10.9 | 141.5 | 38.4 | 121.3 | 9.0 |
| 16 | 18 | 11.8 | 20.2 | 2.5 | 15.6 | 2.1 | 3.6 |

Analysis restricted to the independence and attendance questions. Models BRD7, BRD9, BRD7(MAR) and BRD9(MAR). Observed data; fit to the observed data; prediction of the full data; completion of the observed data, using the model fit.

All models BRD6–9 being of the MNAR type, it is tempting to conclude that all evidence points to MNAR as the most plausible missing-data mechanism. This notwithstanding, one cannot even so much as formally exclude MAR. Indeed, Molenberghs et al. (2007) have shown that, for every MNAR model, there is an associated MAR counterpart that reproduces the fit to the observed data but predicts the unobserved data given the observed ones in a fashion consistent with MAR. These counterparts can be seen as versions of their parent model, constrained to retain fit but force prediction to be MAR. The corresponding estimates for the proportion $\theta$ in favor of independence are presented in the last column of Table 2. Let us zoom in on BRD1, 2, 7 and 9. Only BRD7 and BRD9 saturate the observed-data degrees of freedom. The incomplete data as observed, as fitted by each of the four models, and as fitted by these four models' MAR counterparts, are displayed in Tables 4 and 5. The fits of models BRD7, BRD9 and their MAR counterparts coincide with the observed data: every model produces

TABLE 5

*The Slovenian public opinion survey*

| | | | | | | | |
|---|---|---|---|---|---|---|---|
| Fit of BRD1 and BRD1(MAR) to incomplete data | | | | | | | |
| 1381.6 | 101.7 | 182.9 | | 179.7 | 18.3 | 136.0 | |
| 24.2 | 41.4 | 8.1 | | | | | |
| Prediction of BRD1 and BRD1(MAR) to complete data | | | | | | | |
| 1381.6 | 101.7 | 170.4 | 12.5 | 176.6 | 13.0 | 121.3 | 9.0 |
| 24.2 | 41.4 | 3.0 | 5.1 | 3.1 | 5.3 | 2.1 | 3.6 |
| Completed data using BRD1≡BRD1(MAR) fit | | | | | | | |
| 1439 | 78 | 148.1 | 10.9 | 141.5 | 38.4 | 121.3 | 9.0 |
| 16 | 16 | 11.9 | 20.1 | 2.5 | 15.6 | 2.1 | 3.6 |
| Fit of BRD2 and BRD2(MAR) to incomplete data | | | | | | | |
| 1402.2 | 108.9 | 159.0 | | 181.2 | 16.8 | 136.0 | |
| 15.6 | 22.3 | 32.0 | | | | | |
| Prediction of BRD2 to complete data | | | | | | | |
| 1402.2 | 108.9 | 147.5 | 11.5 | 179.2 | 13.9 | 105.0 | 8.2 |
| 15.6 | 22.3 | 13.2 | 18.8 | 2.0 | 2.9 | 9.4 | 13.4 |
| Prediction of BRD2(MAR) to complete data | | | | | | | |
| 1402.2 | 108.9 | 147.7 | 11.3 | 177.9 | 12.5 | 121.2 | 9.3 |
| 15.6 | 22.3 | 13.3 | 18.7 | 3.3 | 4.3 | 2.3 | 3.2 |
| Completed data using BRD2 fit | | | | | | | |
| 1439 | 78 | 147.5 | 11.5 | 142.4 | 44.7 | 105.0 | 8.2 |
| 16 | 16 | 13.2 | 18.8 | 1.6 | 9.3 | 9.4 | 13.4 |
| Completed data using BRD2(MAR) fit | | | | | | | |
| 1439 | 78 | 147.7 | 11.3 | 141.4 | 40.2 | 121.2 | 9.3 |
| 16 | 16 | 13.3 | 18.7 | 2.6 | 13.8 | 2.3 | 3.2 |

Analysis restricted to the independence and attendance questions. Models BRD1, BRD2, BRD1(MAR) and BRD2(MAR). Fit to the observed data; prediction of the full data; completion of the observed data, using the model fit.

exactly the same fit as does its MAR counterpart. Since BRD1 is MCAR and hence MAR, it is the only one coinciding with its MAR counterpart. Further, while BRD7 and BRD9 produce a different prediction of the complete data, BRD7(MAR) and BRD9(MAR) coincide, owing to saturation. An observation for model assessment and selection is that the five models BRD6, BRD7, BRD8, BRD9 and BRD6(MAR) ≡ BRD7(MAR) ≡ BRD8(MAR) ≡ BRD9 at the same time saturate the observed-data degrees of freedom and exhibit a dramatically different prediction of the full data, and hence for $\theta$: 0.741, 0.764, 0.867, 0.819 and 0.892.

Additional problems can occur, such as predicted complete tables with negative counts, as reported by BRD, Molenberghs et al. (1999) and Molenberghs and Kenward (2007).

## 4. MODEL SELECTION AND ASSESSMENT WITH INCOMPLETE DATA

The five issues laid out at the start of Section 3 and illustrated using both examples, essentially originate from the fact that, when fitting models to incomplete data, one needs to manage two aspects rather than a single one, as schematically represented in Figure 2: the contrast between data and model is supplemented with a second contrast between their complete and incomplete versions.

Ideally, we would want to consider the situation depicted in Figure 2(b), where the comparison is fully made at the complete level. Since the complete data are, by definition, beyond reach, it is tempting but dangerous to settle for the situation in Figure 2(c). This would happen when we would conclude Model 1 fit poorly to the orthodontic growth data, as elucidated by Figure 1(b). Such a conclusion would ignore that the model fit is actually a prediction at the complete-data level, that is, 16 boys and 11 girls, rather than the observed 11 boys and 7 girls, at the age of 10. In other words, one would fail to consider the model's prediction *conditional* on what is observed in the data. Thus, a fair model assessment should be confined to the situations laid out in Figure 2(b) and (d) only. We will start out by the simpler (d) and then return to (b).

Assessing whether Model 1 fits the incomplete version of the growth dataset well can be done by comparing the observed means at the age of 10 to their model fit. This implies we have to confine model fit to those children actually observed at the age of 10.

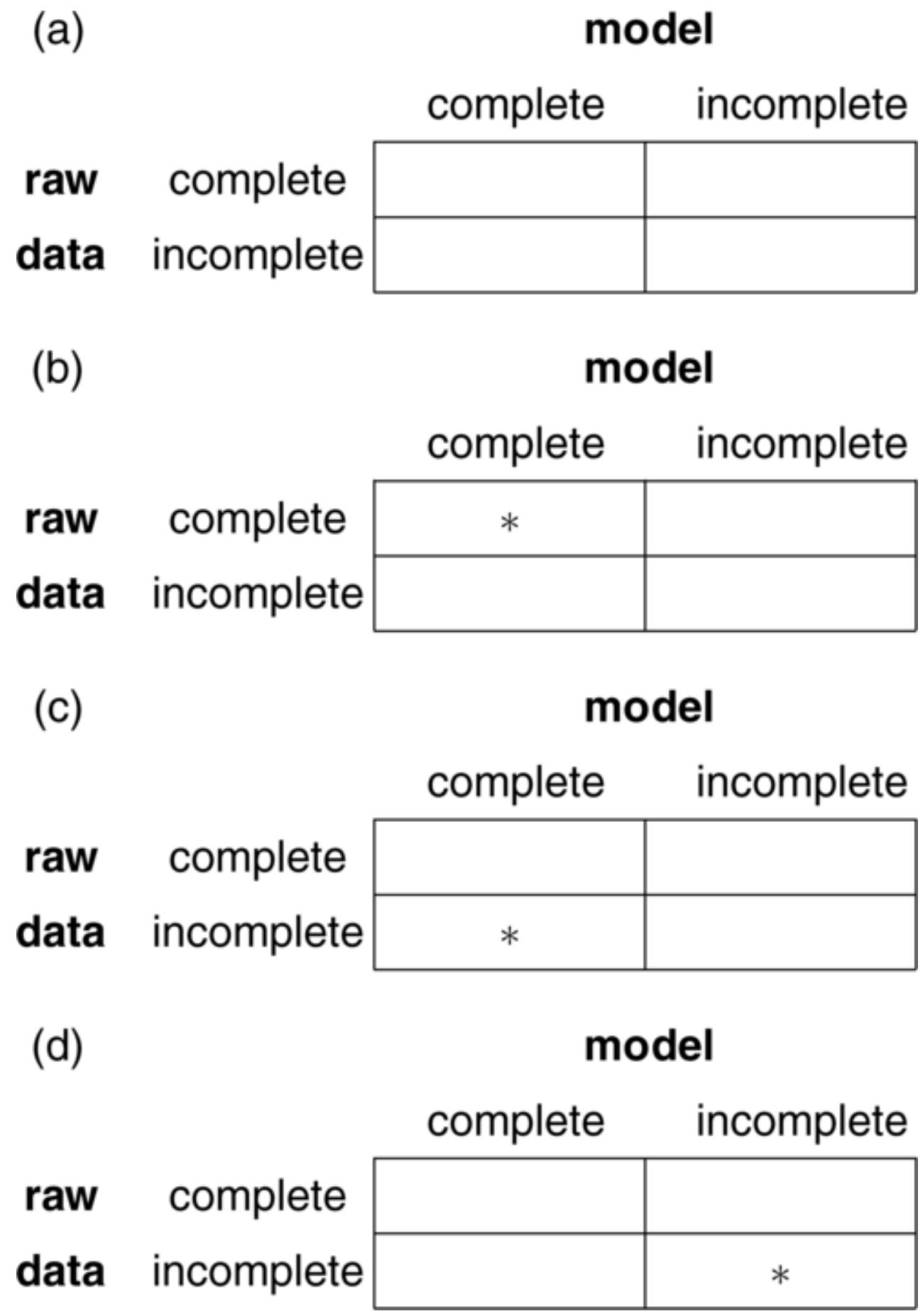

FIG. 2. *Model assessment when data are incomplete.* (a) *Two dimensions in model (assessment) exercise when data are incomplete.* (b) *Ideal situation.* (c) *Dangerous situation, bound to happen in practice.* (d) *Comparison of data and model at coarsened, observable level.*

Turning to the analysis of the SPO, the principle behind Figure 2(d) would lead to the conclusion that the five models BRD6, BRD7, BRD8, BRD9 and BRD6(MAR) ≡ BRD7(MAR) ≡ BRD8(MAR) ≡ BRD9 perfectly fit the observed data. As we stated earlier, though, the models are drastically different in their complete-data level prediction (Table 4) and the corresponding estimates of the proportion in favor of independence, which ranges over [0.74; 0.89]. This points to the need for supplementing model assessment, even when done in the preferable situation of Figure 2(d), with a form of sensitivity analysis.

In conclusion, there are *two* important aspects in selection and assessment when data are incomplete. First, the model needs to fit the *observed* data well. This aspect alone is already quite a bit more complicated than in the complete/balanced case, as shown in Section 3. We will expand on this first aspect in Section 4.1. Second, sensitivity analysis is advisable to assess in how far the model selected and conclusions reached are sensitive to the explicit or implicit assumptions a model makes about the incomplete data, given the observed ones, because such assumptions typically have an impact on the inferences of interest. This aspect is elaborated upon in Section 4.2.

## 4.1 Model Fit to Observed Data

As stated before, model fit to the observed data can be done either by means of what we will label Scenario I, as laid out in Figure 2(b), or by means of Scenario II of Figure 2(d).

Indeed, one of the dangers associated with not considering these scenarios can be clearly illustrated using the orthodontic growth data. Let us take Model 1. When the OLS fit is considered, only valid under MCAR, one would conclude there is a perfect fit to the observed means, also at the age of 10. The estimate from ML would apparently show a discrepancy, since the observed mean refers to a reduced sample size while the fitted mean, similar to (2), is based on the entire design. Thus, it is tempting but incorrect to operate under the scenario of Figure 2(c).

Under Scenario I, for the SPO data, we conclude BRD6–9 or their MAR counterpart fit perfectly. There is nothing wrong with such a conclusion, as long as we realize *there is more than one model* with this very same property, while at the same time they lead to different substantive conclusions. If one would have started with a single one from among these models without considering any of the others, there is a real danger when the conclusions are based on that particular model only. For example, if one would so choose BRD9, the conclusion would be that $\widehat{\theta} = 0.867$ with 95% confidence interval $[0.851; 0.884]$. Ignoring the other perfectly fitting models does not make sense, unless there are very strong substantive reasons to do so.

These considerations suggest that the fit of a model to an incomplete set of data requires caution and perhaps extension and/or modification of the classical model assessment paradigms. In particular, it is of interest to consider assessment under Scenario II.

Gelman et al. (2005) proposed a Scenario II method. The essence of their approach, belonging to the family of so-called posterior predictive model checking, is as follows. First, a model, saturated or nonsaturated, is fitted to the observed data. Under the fitted model, and assuming ignorable missingness, datasets simulated from the fitted model should "look similar" to the actual data. Therefore, multiple sets of data are sampled from the fitted model, and compared to the dataset at hand. Because what one actually observes consists of, not only the actually observed outcome data, but also realizations of the missingness process, comparison with the simulated data would also require simulation from, hence full specification of, the missingness process. This added complexity is avoided by augmenting the observed outcomes with imputations drawn from the fitted model, conditional on the observed responses, and by comparing the so-obtained completed dataset with the multiple versions of simulated complete datasets. Such a comparison will usually be based on relevant summary characteristics such as time-specific averages or standard deviations. As suggested by Gelman et al. (2005), this so-called data-augmentation step could be done multiple times, along multiple-imputation ideas from Rubin (1987). However, in cases with a limited amount of missing observations, the between-imputation variability will be far less important than the variability observed between multiple simulated datasets. This is in contrast to other contexts to which the technique of Gelman et al. (2005) has been applied, for example, situations where latent unobservable variables are treated as "missing."

Let us first apply the method to the orthodontic growth data. The first model considered assumes a saturated mean structure, as in Model 1, with a compound-symmetric covariance structure (Model 1a). Twenty datasets are simulated from the fitted model, and time-specific sample averages are compared to the averages obtained from augmenting the observed data based on the fitted model. The results are shown in the top panel of Figure 3. The sample average at age 10, for the girls, is relatively low compared to what would be expected under the fitted model. Since the mean structure is saturated, this may indicate lack of fit of the covariance structure. We therefore extend the model by allowing for sex-specific covariance structures (Model 1b). The results under this new model are presented in the bottom panel of Figure 3. The observed data are now less extreme compared to what is expected under the fitted model. Formal comparison of the two models, based on a likelihood ratio test, indeed rejects the first model in favor of the second one ($p = 0.0003$), with much more between-subject variability for the girls than for the boys, while the opposite is true for the within-subject variability.

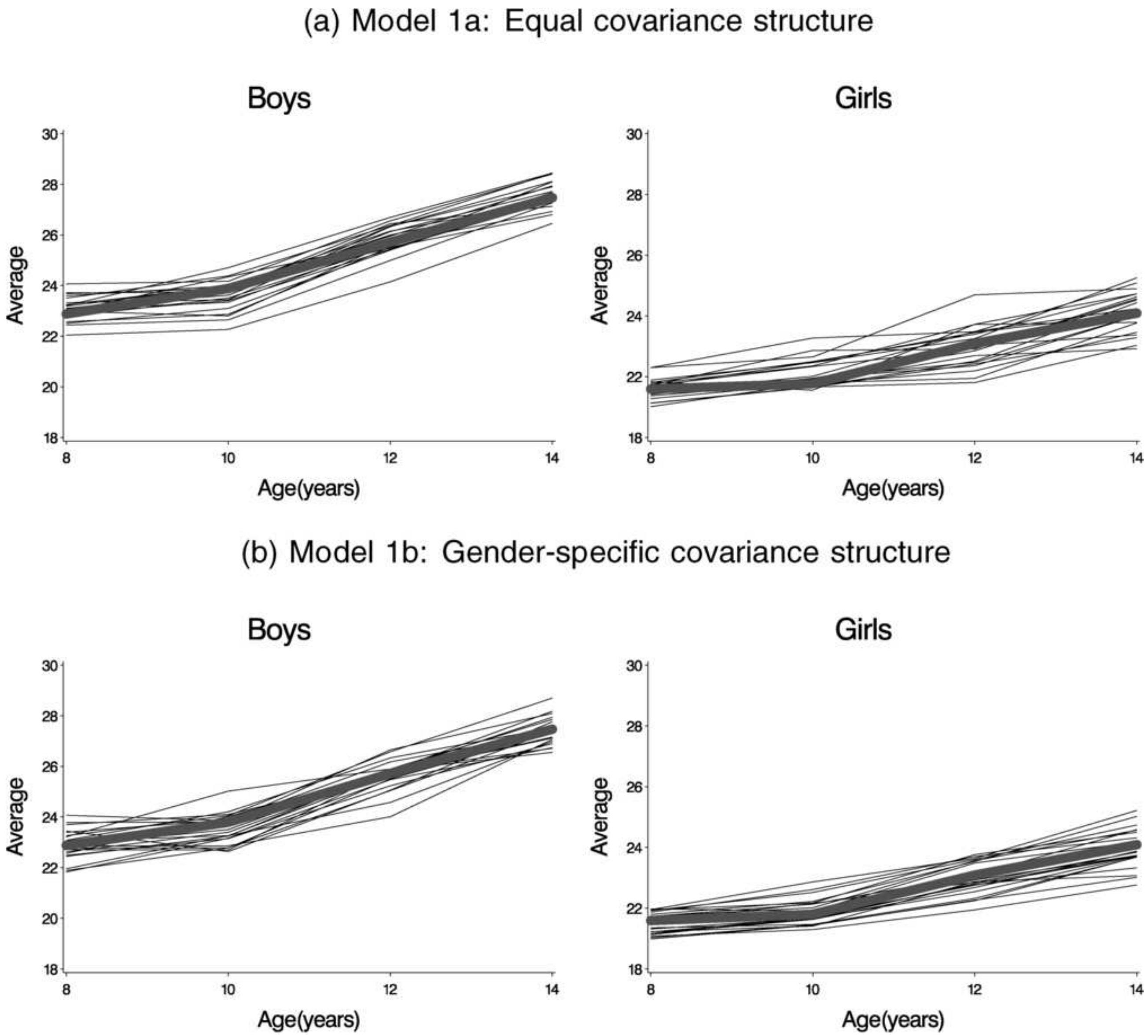

FIG. 3. *The orthodontic growth data. Sample averages for the augmented data (bold line type), compared to sample averages from 20 simulated datasets, based on the method of Gelman et al. (2005). Both models assume a saturated mean structure and compound symmetric covariance. Model 1a assumes the same covariance structure for boys and girls, while Model 1b allows gender-specific covariances.*

Let us now turn to the SPO data. In a contingency table case, the above approach can be simplified to comparing the model prediction of the complete data, such as presented in Tables 4 and 5, with their counterpart obtained from extending the observed, incomplete data to their complete counterpart by means of the fitted model. Here, we have to distinguish between saturated and nonsaturated models. For saturated models, such as BRD6–9 and their MAR counterparts, this is simply the same table as the model's prediction of the full data and again, all models are seen to fit perfectly. Of course, this statement needs further qualification. It still merely means that these models fit the *incomplete* data perfectly, while each one of them tells a different, unverifiable story about the unobserved data given the observed ones. In contrast, for the nonsaturated models, such as BRD1–5 and their MAR counterparts, a so-completed table is different from the predicted one. To illustrate this, the completed tables are presented in Tables 4 (MAR7 and MAR9) and 5 (MAR1 and MAR2).

A number of noteworthy observations can be made. First, BRD1≡BRD1(MAR) exhibits the poorest fit (i.e., the largest discrepancies between this completed table and the model fit), with an intermediate quality fit for a model with 7 degrees of freedom, such as BRD2, and a perfect fit for BRD7, BRD9, and their MAR counterparts. Second, compare the data completed using BRD1 (Table 5) to its prediction of BRD1: the data for the group of completers are evidently equal to the original data (Table 4) since here no completion is necessary; the complete data for the subjects without observations are entirely equal to the model fit, since here there are no data to start from; the complete data for the two partially classified tables take a position in between and hence are not exactly equal to the model prediction. Third, note that the above statement is in need of amendment for BRD2 and BRD2(MAR). Now, the first subtable of partially classified subjects exhibits an exact match between completed data and model prediction, while this is not true for the second subtable. The reason is that BRD2 allows missingness on the second question to depend on the first one, leading to saturation of the first subtable, whereas missingness on the first question is independent of one's opinion on either question.

While the method is elegant and gives us a handle regarding the quality of the model fit to the incomplete data, while contemplating the completed data and the full model prediction, the method is unable to distinguish between the saturated models BRD6–9 and the MAR counterpart, as any method would. This naturally begs the question as to what other methods can be used. In full generality, the literature on model assessment, goodness-of-fit, and diagnostic tools is vast. A relatively early, encompassing reference is D'Agostino and Stephens (1986). Work devoted to the case of longitudinal data was done by Beckman, Nachtsheim and Cook (1987) and Lesaffre and Verbeke (1998). These authors used global and local influence analysis, techniques nicely reviewed in Chatterjee and Hadi (1988), and useful also for sensitivity analysis (see next section). For the specific case of categorical outcomes, work has been done in the context of logistic regression (Hosmer and Lemeshow, 1989) and conventional contingency table analysis (Agresti, 2002), among others. The problem is that many methods are difficult to apply and/or misleading when data are incomplete, thus reducing the analyst's options in this setting. This was the rationale for Gelman's method, as mentioned earlier. An important exception is the family of techniques for contingency tables, where such simple and well-known tools as the likelihood ratio and Pearson's $\chi^2$ goodness-of-fit can be used, at the condition they are applied to the observable cells, of course. Let us apply these to the SPO data. The likelihood ratio test statistics for BRD1, BRD2, BRD7 and BRD9 are 128.46, 72.74, 0 and 0, respectively, on 2, 1, 0 and 0 degrees of freedom, respectively. The corresponding Pearson $\chi^2$ values are 107.9, 50.9, 0 and 0, respectively. This simply confirms our earlier conclusion that BRD1 and BRD2 fit extremely poorly, while the fit of BRD7 and BRD9 is perfect. Of course, this confirmation still does not allow us to make a statement as to the latter models' quality in terms of predicting the unobserved portion of the data, a phenomenon pointing to the need for sensitivity analysis, a topic taken up next.

## 4.2 Sensitivity Analysis

In the previous section, we have seen how one can proceed to assess model fit, either under Scenario I or using Scenario II. It is important to reiterate this comprises the fit to the observed data only, and strictly makes no statement about the model in as far as it describes, or predicts, the unobserved given the observed data. To address the latter issue, a variety of sensitivity analysis routes have been proposed. For our purposes, one could informally define a sensitivity analysis as a way of exploring the impact of a model and/or selected observations on the inferences made when data are incomplete. However, the concept of sensitivity analysis is both older and broader. In Section 4.2.1, we will provide a brief perspective on this vast field, to return to the growth and SPO studies in Section 4.2.2.

4.2.1 *A perspective on sensitivity analysis.* Sensitivity analysis, generically defined as assessment of how scientific conclusions depend on model assumptions, influential observations and subjects, and the like, has a long history in statistics. Early instances include Cornfield's work in the context of causal inference (Holland, 1986) and the study of the independent censoring assumption's impact in time-to-event analyses, to which a large part of a joint U.S. Air Force, National Cancer Institute and Florida State University sponsored conference was devoted (Proschan and Serfling, 1974, and several contributions therein, in particular by Fisher and Kanarek).

A different strand is formed by input/output sensitivity in industrial applications (Haug, Choi and Komkov, 1986).

Even when confining attention to the field of incomplete data, research is vast and disparate. This is not a negative point: rather it reflects broad awareness of the need for such sensitivity analysis. Earlier work on incomplete data was virtually exclusively focused on the formulation of ever more complex models. Both the pattern-mixture model framework (Little, 1993, 1994a) and the shared-parameter framework (Wu and Carroll, 1988; Wu and Bailey, 1988, 1989) have provided useful vehicles for model formulation. In a pattern-mixture model, the outcome distribution is modeled conditional on the observed response pattern, as opposed to the selection-model framework, used throughout this manuscript, where the unconditional outcome distribution is the centerpiece, sometimes supplemented with a model describing the nonresponse process, given the outcomes. In a shared-parameter model, the outcome and nonresponse processes are considered independent, *given* a set of common latent variables or random effects, which are assumed to drive both processes simultaneously. A particularly versatile research line is geared toward the formulation of semiparametric approaches (Robins, Rotnitzky and Zhao, 1994; Scharfstein, Rotnitzky and Robins, 1999). Whereas in the parametric context one is often interested in quantifying the impact of model assumptions, the semiparametric and nonparametric modelers aim at formulating models that have a high level of robustness against the impact of the missing-data mechanism. A number of authors have aimed at quantifying the impact of one or a few observations on the substantive and missing data mechanism related conclusions (Verbeke et al. 2001; Copas and Li, 1997; Troxel, Harrington and Lipsitz, 1998).

A number of early references pointing to the aforementioned sensitivities and responses thereto include Rosenbaum and Rubin (1983), Nordheim (1984), Little (1994b), Rubin (1994), Laird (1994), Vach and Blettner (1995), Fitzmaurice, Molenberghs, and Lipsitz (1995), Molenberghs et al. (1999), Kenward (1998) and Kenward and Molenberghs (1998). Rosenbaum and Rubin (1983) is a pivotal reference for its propensity-scores basis, a technique useful with incomplete data and beyond. A propensity score is, roughly, the probability of an observation being missing or an indication thereof. The method has been used as a basis for missing-data developments in general and sensitivity analysis in particular. For example, it is strongly connected to more recent inverse probability weighting methods, as well as to certain forms of multiple imputation (Rubin, 1987).

Apart from considering pattern-mixture models (PMM) for their own sake, they have been considered by way of a useful contrast to selection models, either (1) to answer the same scientific question, such as marginal treatment effect or time evolution, based on these two rather different modeling strategies, or (2) to gain additional insight by supplementing the selection model results with those from a PMM approach. Pattern-mixture models also have a special role in some multiple-imputation based sensitivity analyses. Examples of PMM applications can be found in Cohen and Cohen (1983), Muthén, Kaplan, and Hollis (1987), Allison (1987), McArdle and Hamagani (1992), Little and Wang (1996), Little and Yau (1996), Hedeker and Gibbons (1997), Hogan and Laird (1997), Ekholm and Skinner (1998), Molenberghs, Michiels and Kenward (1998), Michiels, Molenberghs and Lipsitz (1999), Verbeke, Lesaffre and Spiessens (2001), Michiels et al. (2002), Thijs et al. (2002) and Rizopoulos, Verbeke and Lesaffre (2007). Whereas the earlier references primarily focus on the use of the framework as such, the later ones [emanate] a gradual shift toward sensitivity analysis applications. Molenberghs et al. (1998) and Kenward, Molenberghs and Thijs (2003) studied the relationship between selection models and PMMs. The earlier paper presents the PMM's counterpart of MAR, whereas the later one states how pattern-mixture models can be constructed such that dropout does not depend on future points in time.

Turning to the shared-parameter (SPM) framework, one of its main advantages is that it can easily handle nonmonotone missingness. Nevertheless, these models are based on very strong parametric assumptions, such as normality of the shared random effect(s). Of course, sensitivities abound in the selection and PMM frameworks as well, but the assumption of unobserved, random or latent effects further compounds the issue. Various authors have considered model extensions. An overview is given by Tsonaka, Verbeke and Lesaffre (2007), who consider shared-parameter models without any parametric assumptions for the shared parameters. A theoretical assessment of the sensitivity with respect to these parametric assumptions is presented in Rizopoulos, Verbeke and Molenberghs (2008).

Beunckens et al. (2007) proposed a so-called *latent-class mixture model*, bringing together features of all three frameworks. Information from the location and evolution of the response profiles, a selection-model concept, and from the dropout patterns, a pattern-mixture idea, is used simultaneously to define latent groups and variables, a shared-parameter feature. This brings several appealing features. First, one uses information in a more symmetric, elegant way. Second, apart from providing a more flexible modeling tool, there is room for use as a sensitivity analysis instrument. Third, a strong advantage over existing methods is the ability to classify subjects into latent groups. If done with due caution, it can enhance substantive knowledge and generate hypotheses. Fourth, while computational burden increases, fitting the proposed method is remarkably stable and acceptable in terms of computation time. Clearly, neither the proposed model nor any other alternative can be seen as a tool to definitively test for MAR versus MNAR, as discussed earlier. This is why the method's use predominantly lies within the sensitivity analysis context. Such a sensitivity analysis is of use both when it modifies the results of a simpler analysis, for further scrutiny, as well as when it confirms these.

As stated earlier, a quite separate, extremely important line of research starts from a semiparametric standpoint, as opposed to the parametric take on the problem that has prevailed throughout this section. Within this paradigm, weighted generalized estimating equations (WGEE), proposed by Robins, Rotnitzky and Zhao (1994) and Robins and Rotnitzky (1995), play a central role. Rather than jointly modeling the outcome and missingness processes, the centerpiece is inverse probability weighting of a subject's contribution, where the weights are specified in terms of factors influencing missingness, such as covariates and observed outcomes. These ideas are developed in Robins, Rotnitzky and Scharfstein (1998) and Scharfstein, Rotnitzky and Robins (1999). Robins, Rotnitzky and Scharfstein (2000) and Rotnitzky et al. (2001) employ this modeling framework to conduct sensitivity analysis. They allow for the dropout mechanism to depend on potentially unobserved outcomes through the specification of a nonidentifiable sensitivity parameter. An important special case for such a sensitivity parameter, $\boldsymbol{\tau}$ say, is $\boldsymbol{\tau} = \mathbf{0}$, which the authors term explainable censoring, which is essentially a sequential version of MAR. Conditional upon $\boldsymbol{\tau}$, key parameters, such as treatment effect, are identifiable. By varying $\boldsymbol{\tau}$, sensitivity can be assessed. As such, there is similarity between this approach and the interval of ignorance concept, touched upon in the second paragraph of the next section. There is a connection with pattern-mixture models too, in the sense that, for subjects with the same observed history until a given time $t-1$, the distribution for those who drop at $t$ for a given cause is related to the distribution of subjects who remain on study at time $t$.

Fortunately, it is often possible in problems of missing data, to bring in assumptions that are external to this study, in the sense of them being untestable from its data, but that are implied by the scientific body of knowledge surrounding the problem. An example is the so-called *exclusion restriction* in certain problems of causal inference. When such assumptions are brought in, the missing-data distribution can become identifiable or, at least, the universe of possibilities may be reduced in size. In particular, such knowledge may provide external evidence against MAR. Key references include Angrist, Imbens and Rubin (1996), Little and Yau (1996) and Frangakis and Rubin (2002). Their work is geared toward both study design and analysis methodology that can integrate such external knowledge.

Thus clearly, the field of sensitivity analysis, for incomplete data and beyond, is both blessed with a long and rich history and vibrantly alive. We will now narrow our focus to a few methods that have particular use in addressing issues raised by the growth and SPO cases.

It is clear that the amount of work in this field is vast. Classifying sensitivity analysis methods by means of a useful taxonomy is easier said than done. One could categorize according to the model family to which they are directed within which they are cast. Alternatively, one can distinguish between context-free techniques and methods that make use of substantive considerations. Some methods make simplifying assumptions and specific choices. For example, a number of sensitivity analysis tools are based upon considering a scalar or low-dimensional sensitivity parameter, often positioned within the original model at one of many possible locations. Such choices are entirely reasonable, and ought to be seen as a pragmatic compromise between the desire to explore sensitivity while keeping the ensuing analysis practically feasible and interpretable.

4.2.2 *Sensitivity analysis for the growth and SPO studies.* Verbeke et al. (2001), Thijs, Molenberghs and Verbeke (2000), Molenberghs et al. (2001), Van Steen et al. (2001) and Jansen et al. (2003) advocated the use of local influence-based methods for sensitivity analysis purposes. Details can be found in Verbeke and Molenberghs (2000), Molenberghs and Verbeke (2005) and Molenberghs and Kenward (2007). The essence of the method is that (i) a subject-specific perturbation is added to the model, for example, by replacing the parameter describing MNAR missingness in the model by Diggle and Kenward (1994) with a subject-specific perturbation:

$$\begin{aligned}&\operatorname{logit}[P(\text{dropout at occasion } j|y_{i,j-1}, y_{i,j})]\\&\quad = \psi_0 + \psi_1 y_{i,j-1} + \omega_i \psi_2 y_{ij},\end{aligned} \tag{4}$$

(ii) then observing that $\omega_i \equiv 0$ corresponds to MAR, and (iii) finally studying the impact of small perturbations of $\omega_i$ around zero. Indeed, a model like (4) is necessary, since for an MNAR model, not only the measurements need to be modeled (e.g., using a linear mixed model); also the dropout mechanism needs to be modeled as a function of the measurements and, in some cases, covariates. Technically, this is done by formally studying the curvature of the likelihood surface. Details can be found in the aforementioned references, as well as in Verbeke and Molenberghs (2000) and Molenberghs and Verbeke (2005). In a variety of examples, the above authors showed that one or a few observations are sometimes able to drive the conclusions about the missing-data mechanism. We applied the method to the orthodontic growth data, assuming either Model 1 or Model 7. The results are qualitatively the same and we present the Model 1 results only. Subjects #3 (girl) and #13, #23 and #27 (boys) come out as very influential. In addition, some influence is seen for #6 and #9 (girls), and #16 (boy). As can be seen from Figure 4, all of these are incomplete, which is different from other applications of the method. Of course, all but one of these are positioned relatively low, and one cannot conclude definitively whether either their incompleteness status or the location of their profile is determining their influence. The influence measure informally described above and denoted by $C_i$ is presented in Figure 5. Even though the $C_i$ measure exhibits very high peaks, removing the highly influential subjects does not alter the substantive conclusions.

Molenberghs, Kenward and Goetghebeur (2001) and Kenward, Goetghebeur and Molenberghs (2001) suggested the use of so-called *regions of ignorance*, combining uncertainty owing to finite sampling with uncertainty resulting from incompleteness. Broadly speaking, they consider overspecified models which then produce nonunique solutions of the likelihood equations. For a single (vector) parameter, the resulting solution is called the interval (region) of ignorance. When uncertainty stemming from finite sampling is added, by superimposing ignorance regions with confidence regions, a wider interval (region) of uncertainty is obtained. A formal basis for such an approach was provided by Vansteelandt et al. (2006). For the SPO data, this comes down to considering models with nine or more degrees of freedom.

The estimated intervals of ignorance and intervals of uncertainty are shown in Table 2. Model 10 is defined as $(\alpha_k, \beta_{jk})$ with

$$\beta_{jk} = \beta_0 + \beta_j + \beta_k, \tag{5}$$

while Model 11 assumes $(\alpha_{jk}, \beta_j)$ and uses

$$\alpha_{jk} = \alpha_0 + \alpha_j + \alpha_k. \tag{6}$$

Finally, Model 12 is defined as $(\alpha_{jk}, \beta_{jk})$, a combination of both (5) and (6). Model 10 shows an interval of ignorance which is very close to [0.741, 0.892], the range produced by the models BRD1–BRD9, while Model 11 is somewhat sharper and just fails to cover the plebiscite value. However, it should be noted that the corresponding intervals of uncertainty contain the true value.

Interestingly, Model 12 virtually coincides with the nonparametric range even though it does not saturate the complete-data degrees of freedom. To do so, not two but in fact seven sensitivity parameters would have to be included. Thus, it appears that a relatively simple sensitivity analysis is sufficient to increase the insight in the information provided by the incomplete data about the proportion of valid YES votes.

## 5. CONCLUDING REMARKS

In this paper, we have illustrated the complexities arising when fitting models to incomplete data. By means of two case studies, the continuous longitudinal orthodontic growth data and the discrete Slovenian Public Opinion Survey data, five generic issues were brought to the forefront: (i) the classical relationship between observed and expected features is convoluted since one observes the data only partially while the model describes all data; (ii) the independence of mean and variance parameters in a (multivariate) normal is lost, implying increased sensitivity, even under MAR; (iii) also the well-known agreement between the (frequentist) OLS and maximum likelihood estimation methods for normal models is lost, as soon as the missing-data mechanism is not of the MCAR type, with related results holding in the nonnormal case; (iv) in a likelihood-based context, deviances and related information criteria cannot be used in the same way as with complete data since they provide no information about a model's prediction of the unobserved data; and, in particular, (v) several models may saturate the observed-data degrees of freedom, while providing a different prediction of the complete data, that is, they only coincide in as far as they describe the observed data; as a consequence, different inferences may result from different saturated models.

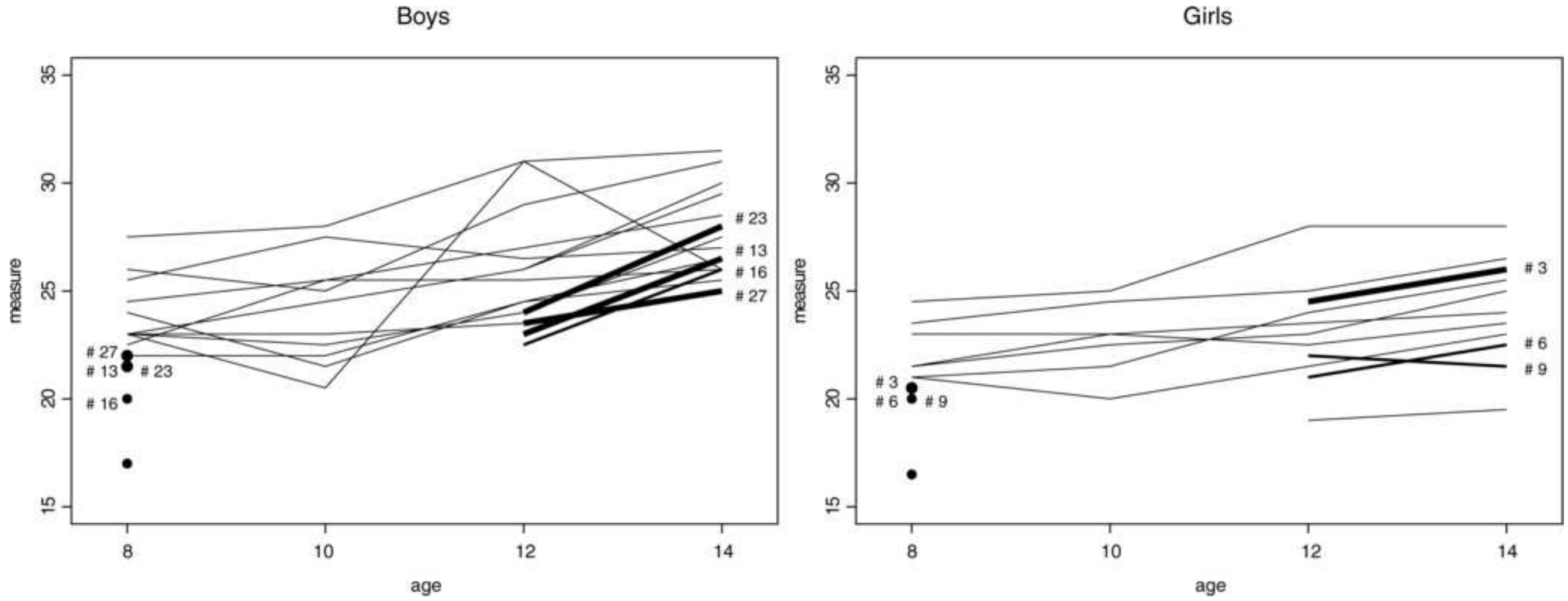

Fig. 4. *The orthodontic growth data. Individual profiles of the incomplete version of the data, with highly and moderately influential subjects highlighted by more and less boldface line type, respectively.*

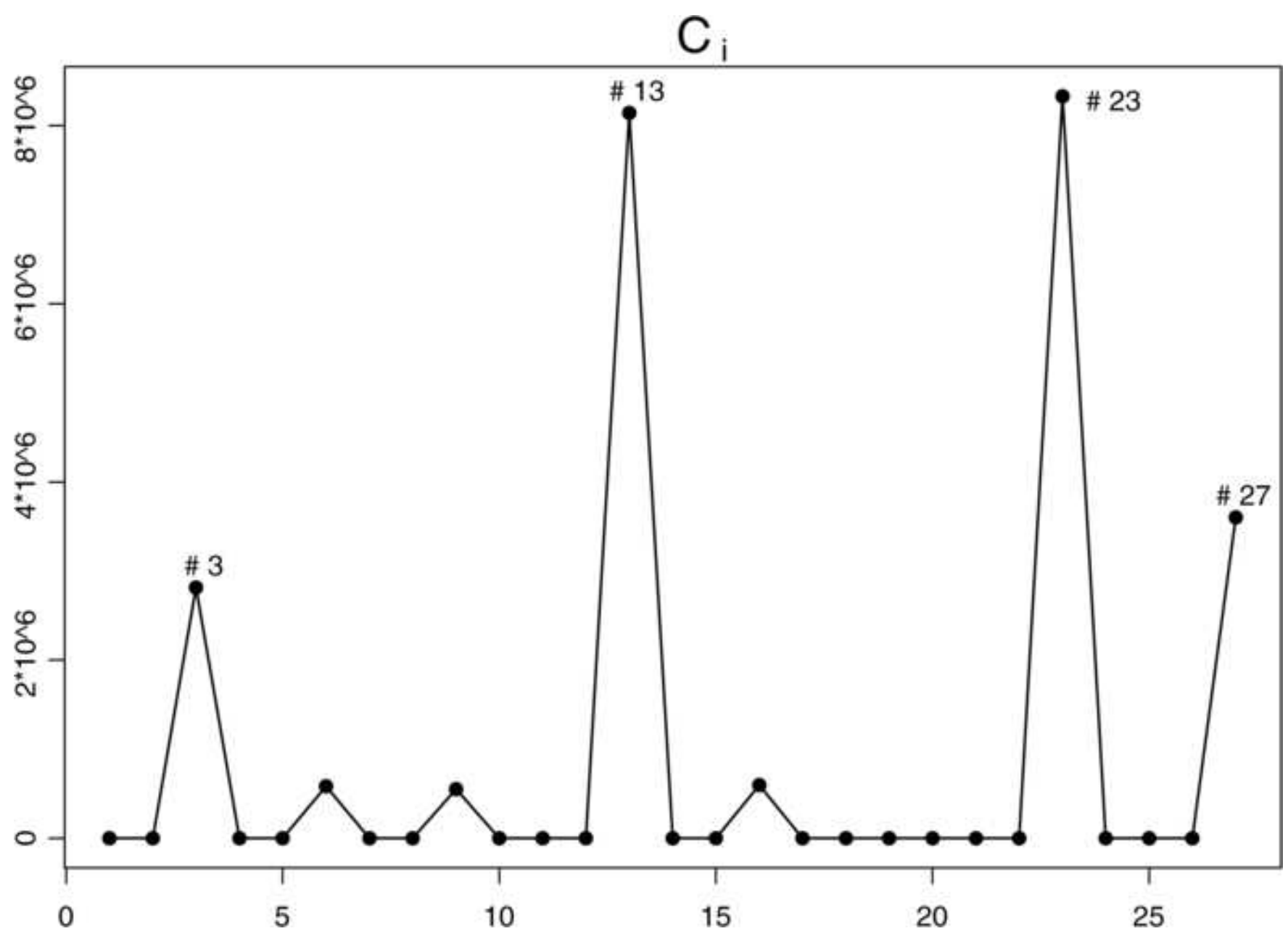

Fig. 5. *The orthodontic growth data. Local influence measures.*

Based on these considerations, it is argued that model assessment should always proceed in two steps. In the first step, the fit of a model to the *observed* data should be assessed carefully, while in the second step the sensitivity of the conclusions to the *unobserved data given the observed data* should be addressed. In the first step, one should ensure that the required assessment be done under one of two allowable scenarios, as represented by Figure 2(b) and (d), thereby carefully avoiding the scenario of Figure 2(c), where the model at the complete-data level is compared to the incomplete data; apples and oranges as it were. The method proposed by Gelman et al. (2005) offers a convenient route to model assessment.

## ACKNOWLEDGMENTS

We gratefully acknowledge support from Belgian IUAP/PAI network "Statistical Techniques and Modeling for Complex Substantive Questions with Complex Data."